\documentclass[12pt,preprint]{aastex}
\slugcomment{V0.4@20111102}

\begin{document}
\title{Discrete Mass Ejections from the Be/X-ray Binary A0535+26/HD~245770}
\author{Jingzhi Yan\altaffilmark{1,2}, Hui Li\altaffilmark{1,2}, and Qingzhong Liu\altaffilmark{1,2} }
\altaffiltext{1}{Purple Mountain Observatory, Chinese Academy of Sciences, Nanjing , China; {\sf
jzyan@pmo.ac.cn,hli@pmo.ac.cn, qzliu@pmo.ac.cn}}
\altaffiltext{2}{Key Laboratory of Dark Matter and Space Astronomy, Chinese Academy of Sciences}

\begin{abstract}
We present the long-term optical spectroscopic observations on the Be/X-ray binary A0535+26 from 1992 to 2010. Combining with the public V-band photometric data, we find that each giant X-ray outburst occurred in a fading phase of the optical brightness. The anti-correlation between the optical brightness and the H$\alpha$ intensity during our 2009 observations indicates a mass ejection event had taken place before the 2009 giant X-ray outburst, which might cause the formation of a low-density region in the inner part of the disk. The similar anti-correlation observed around 1996 September indicates the occurrence of the mass ejection, which might trigger the subsequent disk loss event in A0535+26.
\end{abstract}
\maketitle
\section{Introduction}
Be/X-ray binary represents a major subclass of high mass X-ray binary in our Galaxy \citep{Liu06}. A neutron star moves in a wide and eccentric orbit around the Be star, which is a non-supergiant rapid-rotating B-type star and ever shows Balmer emission lines at least once in its life \citep{Porter03}. There are two different disks in Be/X-ray binary: a circumstellar disk around Be star and an accretion disk around neutron star (see \citet{Reig11} and the reference therein).

A0535+26 was found by Ariel V in 1975 during a giant X-ray outburst \citep{Rosenberg75}. Its optical counterpart was identified as the Be star HD 245770 with a classification of O9.7 IIIe \citep{Giovannelli92}. A neutron star with a 103-s spin period \citep{Caballero07} moves around the primary star in an eccentric orbit (e = 0.47 $\pm$ 0.02) \citep{Finger94}. A 111.1-day orbital period was found from the periodic Type I X-ray outbursts around the periastron passage of the neutron star. Since the first giant outburst in 1975, another six giant X-ray outbursts were observed in 1975 April, 1980 October \citep{Nagase82}, 1983 June \citep{Sembay90}, 1989 March \citep{Makino89}, 1994 February \citep{Finger94}, 2005 May \citep{Tueller05,Coe06,Caballero08}, and 2009 December \citep{Wilson-Hodge09,Reynolds2010}, respectively.

The long-term variability of optical and IR photometry was discussed by \citet{Hao96}, \citet{Clark98}, \citet{Clark99}, \citet{Haigh99}, \citet{Lyuty00}, and \citet{Haigh04}. The variability of H$\alpha$ emission line is related with the physical changes in the circumstellar disk around Be star. A disk loss event took place in the system in 1998 and the H$\alpha$ line changed from emission to absorption \citep{Haigh04,Grundstrom07}. Recent short-term and long-term optical spectral variabilities were reported by \citet{Moritani2010}.

In this paper, we present our optical spectroscopic and photometric observations on A0535+26. Combining with the public data of the V-band magnitude and the H$\alpha$ equivalent widths (EWs), we discuss the optical variability of the system and unveil the nature of X-ray outbursts.

\section{Observations}\label{SecObs}

We obtained optical spectra of HD 245770  with the 2.16~m telescope at Xinglong Station of National Astronomical Observatories, China (NAOC), from 1992 to 2010.
The optical spectroscopy with an intermediate resolution of 1.22~{\AA}\,pixel$^{-1}$ was made with a CCD grating spectrograph at the Cassegrain focus of the telescope. We took both blue and red spectra covering the wavelength ranges 4300--5500 and 5500--6700\,{\AA}, respectively, at different times. All spectra were reduced with the IRAF\footnote{IRAF is distributed by NOAO, which is operated by the Association of Universities for Research in Astronomy, Inc., under cooperation with the National Science Foundation.} package. The data were bias-subtracted and flat-field corrected, and the cosmic rays were removed. Helium-argon spectra were taken in order to obtain the pixel-wavelength relation. To improve this relation, we also used the diffuse interstellar bands 6614 and 6379~{\AA} observed in the spectra.
The short-term and long-term variability of the emission line profile will be discussed in another paper \citep{Liu11}. In this paper we only concentrate on the long-term evolution of the intensity of the H$\alpha$ line in the spectra of A0535+26.  All the spectra have been normalized to neighboring continuum. The EW of the H$\alpha$ line has been measured selecting a continuum point on each side of the line and integrating the flux relative to the straight line between the two points using the procedures available in IRAF. The measurements were repeated five times for each spectrum and the error estimated from the distribution of obtained values. The EW(H$\alpha$) typical error is within 3\%. This error arises due to the subjective selection of the continuum.

On 2010 October 13, we performed the systematic Johnson-Cousin $UBVRI$ photometric observations on A0535+26 with the 80cm Tsinghua-NAOC Telescope (TNT) at Xinglong Station of NAOC. TNT is an equatorial-mounted Cassegrain system with a focal of $f/10$, made by AstroOptik, funded by Tsinghua University in 2002 and jointly operated with NAOC. The telescope is equipped with the Princeton Instrument 1340$\times$1300 thin back-illuminated CCD. The photometric data reduction was performed using standard routines and aperture photometry packages in IRAF, including bias subtraction and flat-field correction. In order to calibrate the instrumental magnitudes of A0535+26, we also observed a sample of Landolt standards in $UBVRI$ bands. The average $UBVRI$ magnitudes of A0535+26 are, 8.6$\pm$0.03, 9.6$\pm$0.03, 9.2$\pm$0.02, 8.9$\pm$0.02 and 8.5$\pm$0.02, respectively.

\section{Analysis}\label{SecAna}

\subsection{Line Variability during Our 2007-2010 observations}
\begin{center}
\begin{figure}
\centering
\includegraphics[bb=64 38 750 533,width=8cm]{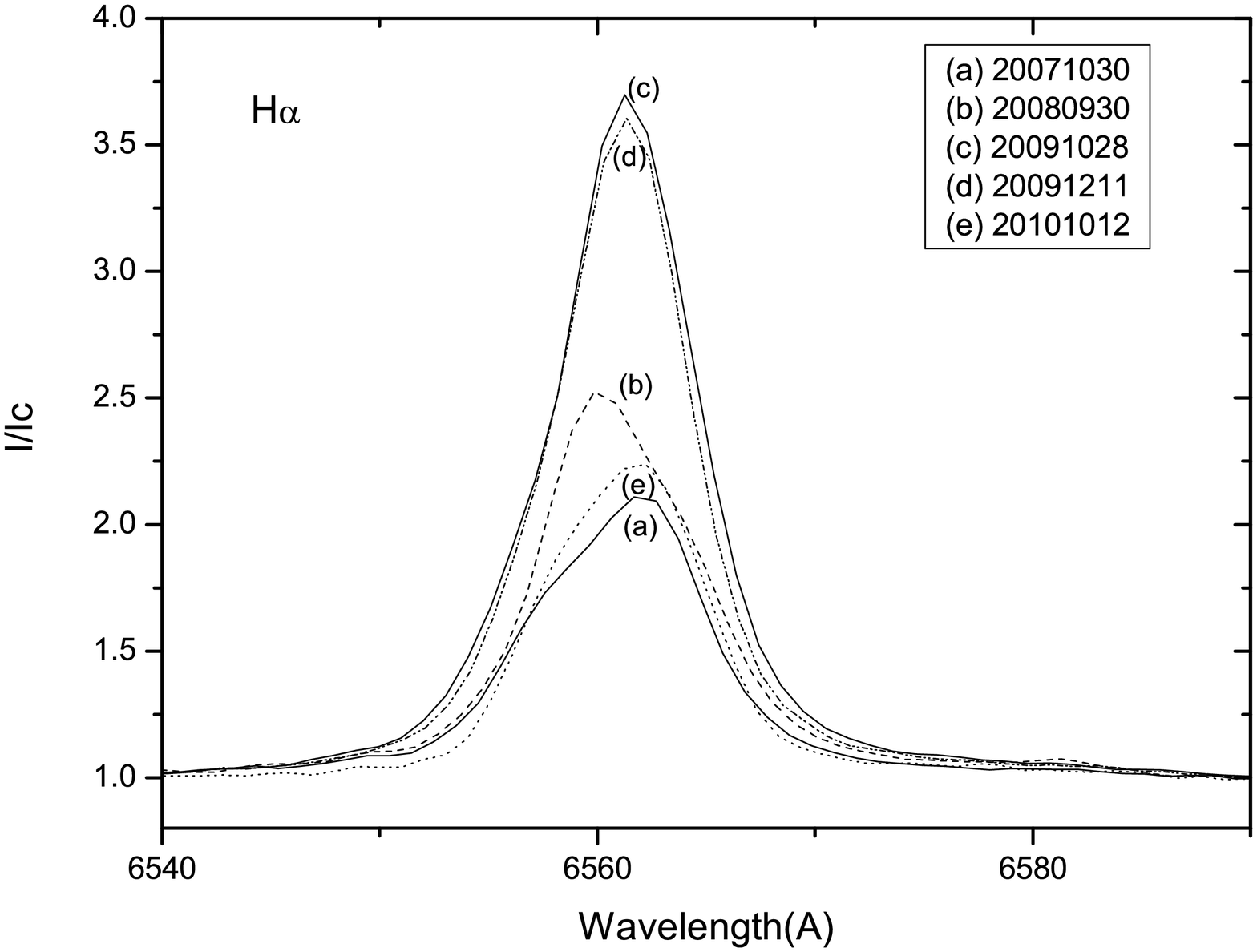}%
\includegraphics[bb=53 33 739 536,width=8cm]{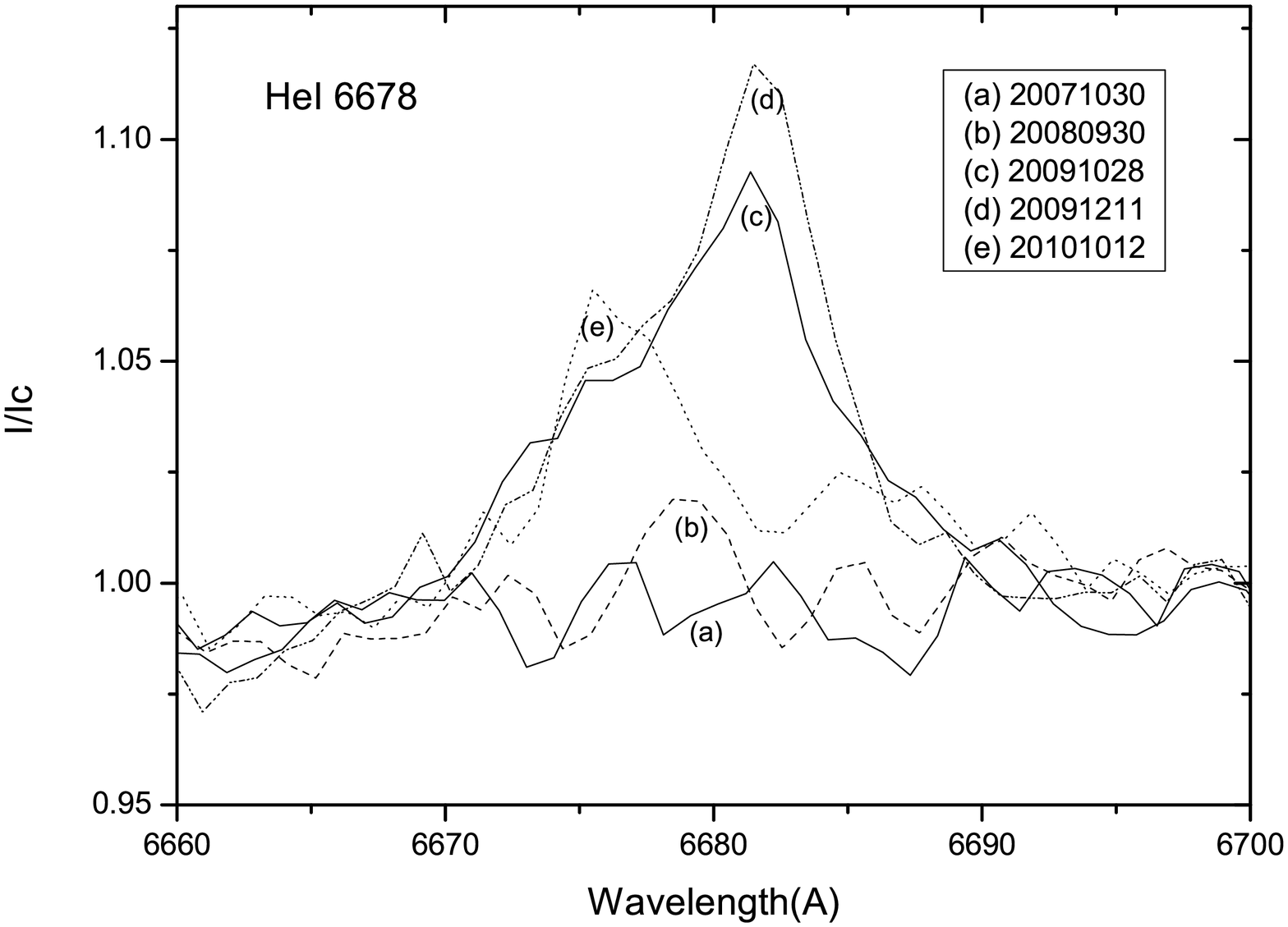}%
\caption{Typical line profiles during our 2007-2010 observations in H$\alpha$ (left) and HeI~$\lambda$~6678 (right) regions, respectively. The insets are the observational dates for each spectrum in a format of YYYYMMDD.}  \label{figure:spectra}
\end{figure}
\end{center}

In order to study the optical line variability before, during, and after the 2009 giant X-ray outburst, we plot the representative spectra of H$\alpha$ and HeI~$\lambda$~6678 during our 2007-2010 observations in Figure~\ref{figure:spectra}. The H$\alpha$ emission line shows an asymmetric single-peaked profile during our 2007 observations. The peak intensity of the H$\alpha$ became stronger during the 2008 observations, which were made just after a small X-ray outburst (see Figure~\ref{figure:hd}). Our 2009 October observations were performed just before the 2009 giant X-ray outburst of A0535+26 and the H$\alpha$ emission showed the strongest intensity of the last twenty years. The H$\alpha$ profiles on 2009 December 11 (MJD~55176), which were observed during the rising phase of the 2009 giant X-ray outburst, showed a subtle change relative to its structure before the outburst. Moreover, the wings of the H$\alpha$ lines in 2009 also became much broader than that in 2007 and 2008. About one year later, the H$\alpha$ emission line returned to its normal level, but its peak intensity was still stronger than that in 2007, while its wings were a bit narrower.

As shown in Figure~\ref{figure:spectra}, the HeI~$\lambda$~6678 line nearly lost its emission feature during our 2007 observations and showed a weak double-peaked profile with R/V$>$1 during the 2008 observations. Like the H$\alpha$ line, HeI~$\lambda$~6678 shows the largest intensity during our 2009 October observations which were obtained before the 2009 giant X-ray outburst. The difference is that the  HeI~$\lambda$~6678 line became stronger during our 2009 December 11 observations and a double-peaked profile was still evident in 2010.

\subsection{Long-term Optical Variability in Different X-ray States} \label{sub_long}

\begin{center}
\begin{figure}
\centering
\includegraphics[height=10cm]{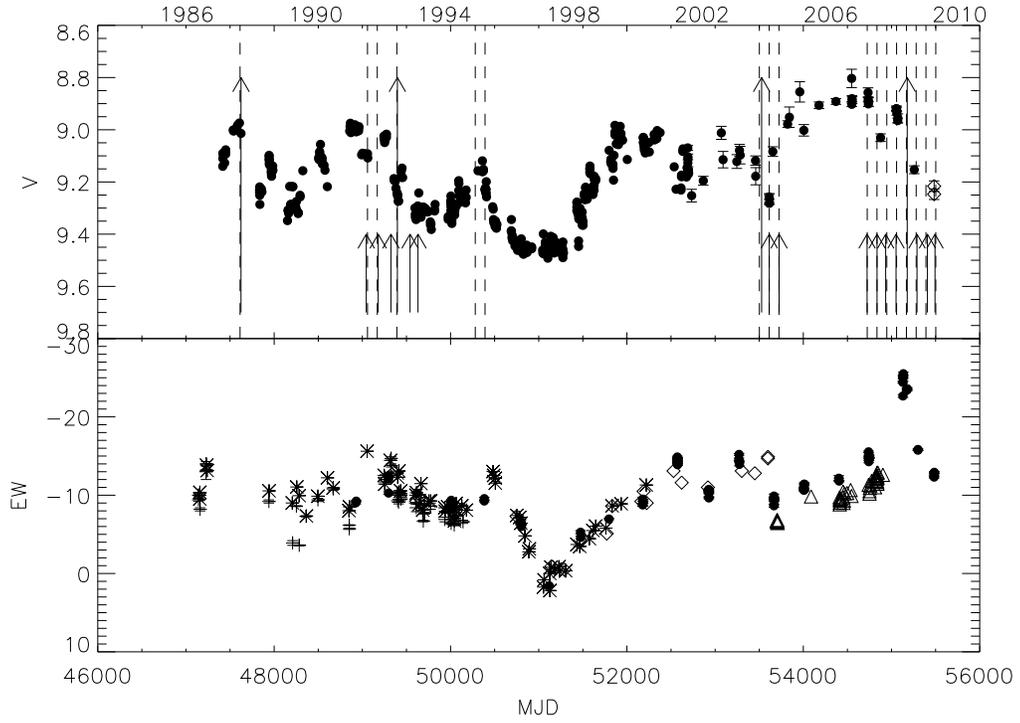}
\caption{Optical observations of A0535+26. The top panel is the V-band magnitudes from literatures and our observations (diamonds); The bottom panel is the EWs of H$\alpha$ emission line from literatures and our observations (solid circles). The long and the short arrows in the top panel correspond to the middle of time for the giant X-ray outbursts and the normal Type I X-ray outbursts, respectively. The dashed lines are the time of the neutron star periastron passages.}
\label{figure:hd}
\end{figure}
\end{center}

\begin{center}
\begin{figure}
\centering
\includegraphics[height=10cm]{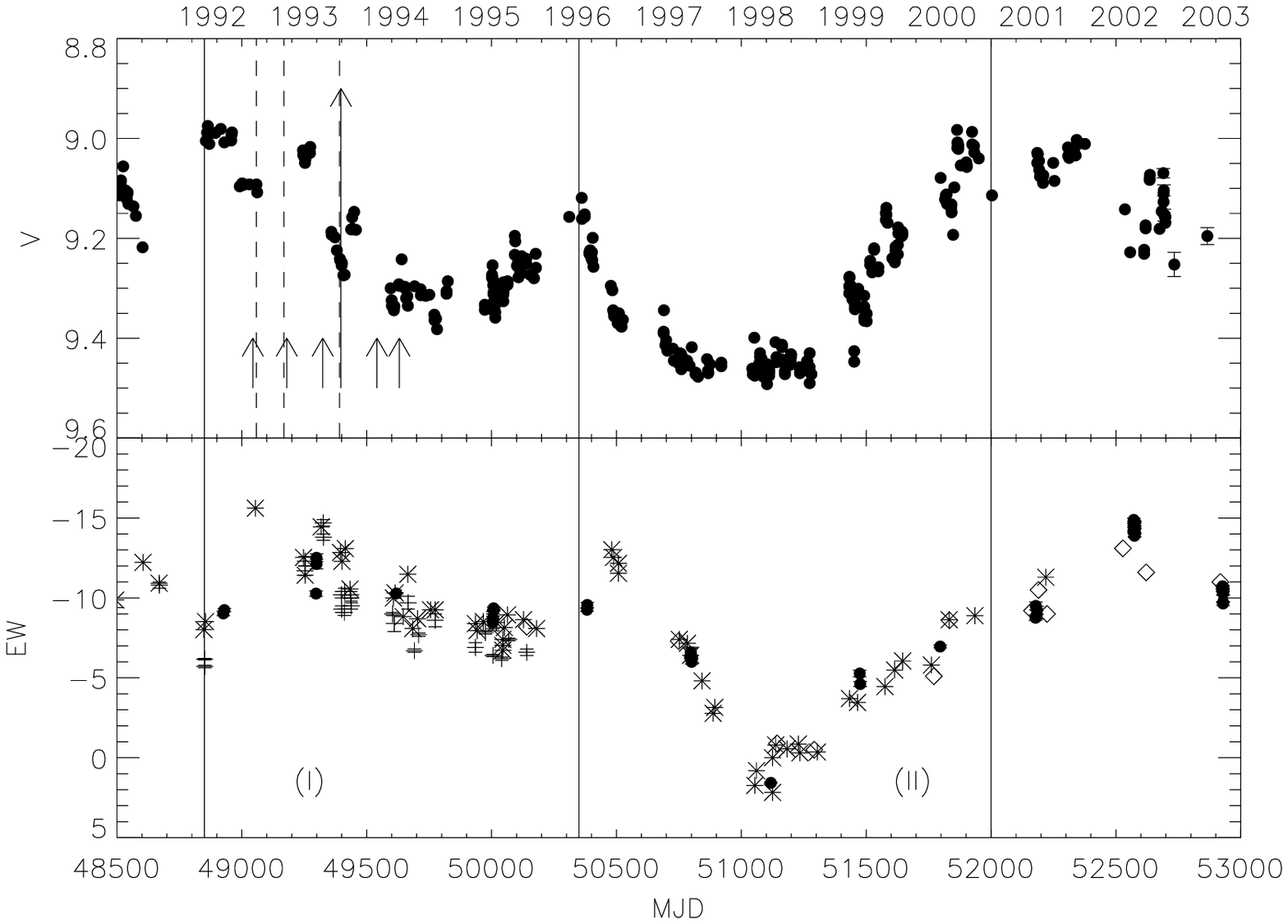}
\caption{The zoomed Figure~\ref{figure:hd} in Cycle I and Cycle II regions.}
\label{figure:hdpart1}
\end{figure}
\end{center}

In this work we also make use of some public V-band photometric data \citep[][and also the INTEGRAL/OMC V-band photometric data, rebinned with a time of 2 days, from the OMC Archive at LAEFF\footnote{http://sdc.laeff.inta.es/omc/index.jsp}]{Lyuty00, Zaitseva05} and the H$\alpha$ EWs data \citep{Clark98,Haigh04,Coe06,Moritani2010}. Our optical photometric and spectroscopic results and all these public data are plotted in Figure~\ref{figure:hd}. The long and short arrows in the top panel of Figure~\ref{figure:hd} represent the middle time of the  giant and normal X-ray outbursts in A0535+26, respectively. The time for the periastron is also plotted with vertical dashed lines in Figure~\ref{figure:hd}, at the ephemeris of, MJD 53613.0+111.1E \citep{Finger96}. In order to better study the relationship between the optical brightness, H$\alpha$ emissivity, and the X-ray activity, we divide Figure~\ref{figure:hd} into Figure~\ref{figure:hdpart1} and Figure~\ref{figure:hdpart2}. Here we focus on the optical variability during the X-ray outbursts and the phase of the disc loss and renewal around the Oe star, that can be divided into four activity cycles.

\begin{itemize}
  \item \textbf{Cycle I} (MJD~48850-50350) A series of X-ray outbursts were observed by BATSE before and after the giant X-ray outburst in 1994 February \citep{Finger96}. \citet{Haigh04} has discussed the multi-wavelength variability of A0535+26 in this cycle. As shown in Figure~\ref{figure:hdpart1}, the first normal X-ray outburst during this cycle took place in an optical brightness decline phase. The V-band apparent magnitude of A0535+26 began to increase around MJD~48850, and reached a magnitude of $M_V$$\sim$ $9.1^m$ around MJD~49050. However, the EWs of H$\alpha$ emission line showed an obvious increase in this time interval. After that, the brightness of the system kept increasing and reached a maximum around MJD~49250. The second X-ray outburst was observed during this optical brightening process. The following normal and giant X-ray outbursts were detected in an abrupt decline phase of the optical emission. Another two normal X-ray outbursts also occurred in the such a phase. The strength of H$\alpha$ kept declining after the giant X-ray outburst. After MJD$\sim$49800, the optical brightness of A0535+26 began to increase and reached a magnitude of $M_V$$\sim$ $9.1^m$ around MJD~50350. In the phase of the optical brightening, the H$\alpha$ EWs did not change considerably.
  \item \textbf{Cycle II} (MJD~50350-52560) During this cycle, A0535+26 underwent the events of the disk loss and renewal. After MJD~50350, the V-band magnitudes increased until MJD~50550, while the strength of H$\alpha$ line became stronger during this time. It is hard to imagine what had happened in the system between MJD~50550 and MJD~50650 due to the bad coverage of the V-band data. But we could deduce from Figure~\ref{figure:hdpart1} that the optical brightness of A0535+26 was still declining after MJD~50650, and this decline phase is different from the former one. After MJD~50550 , the strength of H$\alpha$ decreased continually. When the apparent magnitude of A0535+26 reached a maximum around MJD~50800, the H$\alpha$ intensity still kept declining and reached a minimum around MJD~51050. Then the optical brightness and the intensity of the H$\alpha$ line kept increasing and a new circumstellar disk was reforming around Oe star. When the V-band brightness reached a stable state around MJD~52000, the intensity of the H$\alpha$ emission line was still in an increasing phase and it reached a maximum around MJD~50350.
 \item \textbf{Cycle III} (MJD~53300-53800) During this cycle, a giant X-ray outburst occurred in May/June 2005 after ten years of inactivity of the Be/X-ray binary A0535+26. Two subsequent normal X-ray outbursts were also observed in the system. As shown in Figure~\ref{figure:hdpart2}, the 2005 giant X-ray outburst took place during the increasing phase of the INTEGRAL/OMC V-band magnitudes and the first normal X-ray outburst occurred at the time when its V-band brightness showed a minimum around MJD~53615. The next normal X-ray outburst was observed in the following optical decline phase. In the increasing phase of the V-band magnitude, the strength of the H$\alpha$ emission line became slightly stronger. It is interesting that the strength of the H$\alpha$ line dropped abruptly when the optical brightness began to increase.
  \item \textbf{Cycle IV} (MJD~54600-55600) Eight continuous normal and one giant X-ray outburst was observed during this cycle. Though the time-resolution of the V-band observations is not very good, one can still see from Figure~\ref{figure:hdpart2} that the first four normal X-ray outbursts occurred during the fading and brightening phases of the optical brightness between MJD~54700 and MJD~55100, which is similar to the first two normal X-ray outbursts occurred in Cycle I (see Figure~\ref{figure:hdpart1}). The 2009 December giant X-ray outburst and subsequent three normal X-ray outbursts were observed during another fading phase of the optical brightness. According to our 2010 optical photometric observations, the optical brightness of A0535+26 was still in a low level. The V-band magnitudes and EWs of H$\alpha$ emission line were in a stable level for a long time before Cycle IV. The strength of H$\alpha$ began to increase when the optical brightness started to fade around MJD~54600, and it reached an unprecedented maximum, with an EW of $\sim$--25$\AA$, during our 2009 October observations, which were obtained just before the 2009 giant X-ray outbursts. Our observations on 2009 December 11 (MJD~55176, during the rising phase of the giant X-ray outburst) indicate that the strength of H$\alpha$ was in a slightly lower level than that in our 2009 October observations (also see Figure~\ref{figure:spectra}). Our 2010 observational results indicate that the H$\alpha$ intensity had been decreased to the level before Cycle IV.

\end{itemize}

\section{Discussions}

\begin{center}
\begin{figure}
\centering
\includegraphics[height=10cm]{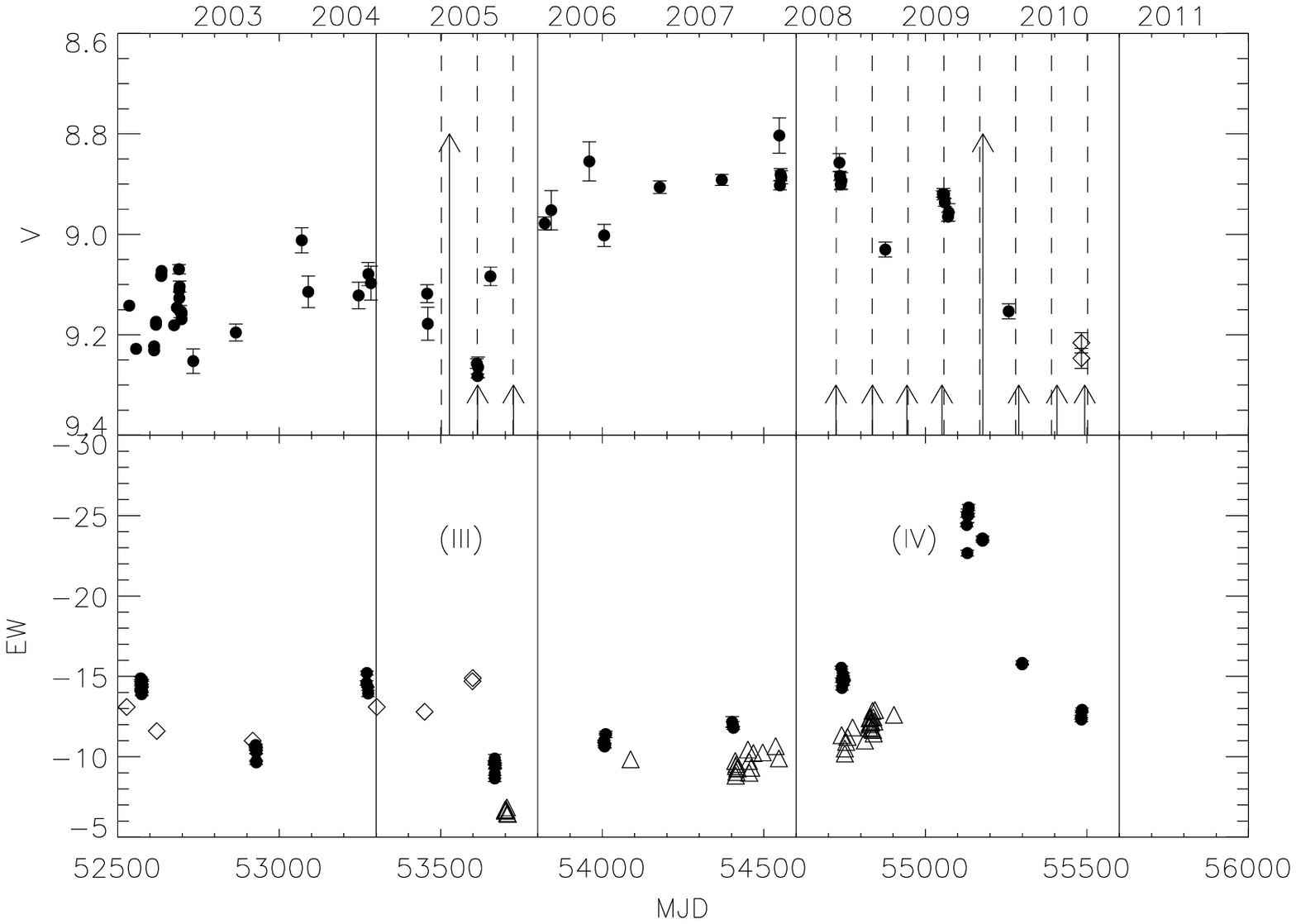}
\caption{The zoomed Figure~\ref{figure:hd} in Cycle III and Cycle IV regions.}
\label{figure:hdpart2}
\end{figure}
\end{center}

\citet{Clark99} found that the X-ray outbursts in Be/X-ray binary A0535+26 usually occurred after a period in which the optical light curve had seen to fade. Our analysis in Subsection~\ref{sub_long} also indicates that the recent 2005 and 2009 giant X-ray outbursts were also observed during a fading phase of the optical brightness (See Figure~\ref{figure:hd}). It is interesting that the intensity of the H$\alpha$ line in the spectra of A0535+26 showed an obvious increase during the fading of the optical brightness in Cycle IV of Figure~\ref{figure:hdpart2}. Similar observational phenomenon was also observed in Cycle I of Figure~\ref{figure:hdpart1}. To explain the anti-correlation between the optical brightness and the H$\alpha$ intensity in A0535+26, we should first know the physical origin of the optical continuum and H$\alpha$ emission in the system.

The optical thick H$\alpha$ emission line in Be star is generally believed to be formed in the entire circumstellar disk \citep{Slettebak92}, while only the innermost part of the disk contributes significantly to the continuum flux. Due to the higher ionization potential energy, the formation region of the HeI~$\lambda$6678 line should be smaller than the nearby continuum region \citep{Stee98}. The strength of the H$\alpha$ line became stronger during our 2009 observations, while its V-band brightness kept fading during that period. The increase of the H$\alpha$ emission might be caused by the decrease of the optical continuum around the H$\alpha$ region. In fact, our optical observations in 2010 indicate that even the optical brightness was still in a fading phase, the strength of the H$\alpha$ emission line had declined to its level in our 2007 observations. Therefore, we suggest that the obvious increase of the H$\alpha$ intensity during our 2009 observations was not only caused by the decrease of the optical continuum in A0535+26. The correlation between the $V$ magnitude and the $(B-V)$ colour in A0535+26 \citep{Clark99} indicates that the variability in the $UBV$ photometric observations was connected with the changes in the emission from the circumstellar disk around the Oe star \citep{Janot-Pacheco87}. Therefore, the only source for the variability of the V-band brightness is the circumstellar disk around the HD~245770. The decrease of the V-band brightness during Cycle IV might indicate that great changes had taken place in the inner region of the disk. The strong H$\alpha$ emission during our 2009 observations might be due to the formation of a larger circumstellar disk before the giant outburst in 2009.
Figure~\ref{figure:spectra} shows that the HeI~$\lambda$6678 became much stronger during our 2009 observations and it still had a strong emission during our 2010 observations, which may be attributed to a denser inner part of the disk formed before the 2009 giant X-ray outburst. It seems to be contradictory that the optical brightness became faint when a dense inner disk was formed, if we assume the optical excess comes from the disk around the Oe star.

A unified model should be assumed to explain all the observed features in A0535+26. Observational results \citep{Rivinius01} and theoretical calculations \citep{Meilland06} suggest that after an outburst in Be star a low-density region seems to develop above the star. About the nature of the outburst is beyond the scope of this paper. \citet{Rivinius01} and \citet{Meilland06} suggested that the outburst might be connected with the increased mass loss or mass ejection from Be star. Some weeks to months after the outburst, the stellar radiation pressure gradually excavates the inner part of the disk and a low-density region could develop around the Be star and slowly grow outwards \citep{Rivinius01}. With the vacuation of the inner disk, the optical continuum emission decreases and an increase in $UBV$ magnitudes will be observed. As shown in Figure~\ref{figure:hdpart2}, the V-band magnitude began to increase at the beginning of Cycle IV. The decrease of the V-band continuum emission indicates that a mass ejection event from Be star had taken place before MJD~54600. After the mass ejection, a low density region was forming around the Be star. Besides the decrease of the optical continuum emission, the vacuation of the inner disk could also cause the inner circumstellar disk changes from optically thick to optically thin, which could increase the volume of the Helium line emission. Therefore, the optical thin line, such as HeI~6678, appeared after the mass ejection, and became stronger when the optical continuum emission was decaying. Moreover, a lot material would be transferred into the disk after the mass ejection event and the outer part of the disk would move outward. A more extended circumstellar disk should correspond to a stronger H$\alpha$ emission. This may explain the detected increasing behavior of H$\alpha$ emission presented in Figure~\ref{figure:hdpart2}.

\begin{center}
\begin{figure}
\centering
\includegraphics[bb=73 35 750 533,width=8cm]{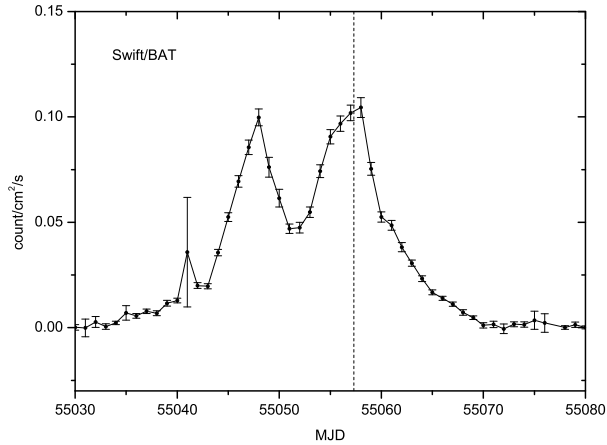}
\caption{The double-peaked X-ray outburst of A0535+26 observed by Swift/BAT in 15-50 keV energy band from MJD~55030 to MJD~55080. The vertical dashed line indicates the time for the periastron passage of the neutron star.}
\label{figure:double}
\end{figure}
\end{center}

The V-band brightness did not show a continuous decline during Cycle IV. The optical brightness of A0535+26 began to decrease around MJD~54600 and showed an increase during the following four normal X-ray outbursts.
The increase of the optical emission indicated that a subsequent outburst had taken place and the cavity region was replenished by the ejected material. With the outward motion of the ejected material, a secondary inner circumstellar disk would be formed around the star and a ring-like structure was developing in the disk. When the disk expanded to a certain extent, the neutron star would come inside the circumstellar envelope of the Be star and accrete matter, becoming a transient source. This may account for the first four normal X-ray outbursts in Cycle IV. A considerable fraction of the angular momentum would be transferred on to the neutron star in the system, which may account for the simultaneous spin-up detected by Fermi/GBM \citep{Camero-Arranz11}. It is interesting that a distinguished double-peaked profile was observed by Swift/BAT in the fourth X-ray outburst during Cycle IV (see Figure~\ref{figure:double}). The profile of the outburst might be connected with the accreting environment of the neutron star. There should be a low-density region in the circumstellar disk when the neutron star was approaching the perisatron point around MJD~55057.3 (the vertical dashed line in Figure~\ref{figure:double}). This low-density region might be the low-density ring structure discussed above. Assuming the circumstellar disk of A0535+26 dissipated on a viscous timescale, we can estimate the viscosity parameter $\alpha$ \citep{Shakura73} using Equation (19) of \citet{Bjorkman05}:
\begin{equation}
\label{euqation1}
t_{diffusion}=(0.2yr/\alpha)*(r/R_*)^{0.5},
\end{equation}
where $t_{diffusion}$ is the diffusion time of the low-density region from the surface of the Oe star to the periastron point and $r$ is the radial distance from the Oe star to the periastron point of the neutron star. We assume that the time for the formation of the low-density region corresponds to the initial moment of the optical brightness decrease, $\sim$MJD~54600. Thus, $t_{diffusion}$ is about 457.3~days. Given that the mass $M_*$ and the radius $R_*$ of the Oe star HD~245770 are $14M_{\odot}$ and $15R_{\odot}$ \citep{Giangrande80}, respectively, and the typical mass for a neutron star is $1.4M_{\odot}$, the value of $r/R_*$ is approximately 8.6, for an orbital period of 111.1~days with an eccentricity of $\sim$0.47. Therefore, the viscosity parameter $\alpha$ is $\sim$0.47 estimated by Equation~\ref{euqation1}. This result indicates that the circumstellar disk around Oe star in A0535+26 will be truncated at the 4:1 resonance radius \citep{Okazaki01}, which will allow enough mass accretion on to the neutron star at the periastron passage to cause the X-ray outburst in the system during Cycle IV.

Very strong H$\alpha$ emission was observed during our 2009 observations, when the V-band brightness showed an obvious decrease, which indicated a strong mass ejection had taken place before our 2009 observations and a lot of material was replenished the circumstellar disk. An extended and dense circumstellar disk caused the strong X-ray outburst during Cycle IV. After the 2009 strong X-ray outburst, the disk around Oe star could also be truncated by the neutron star in A0535+26 and the strength of the H$\alpha$ emission line showed a rapid decline. One year later, the emission of H$\alpha$ line had returned to its level in the quiescent state.

Figure~\ref{figure:hdpart1} shows that the V-band light curve during Cycle I has the same variability as that during Cycle IV. The decrease of the optical brightness during the first stage of Cycle I might also be caused by the discrete mass ejection from the Oe star. The second abrupt decrease of the optical brightness indicated a lot of material was filled in the disk, which caused the 1994 giant X-ray outburst. The anti-correlation between the optical brightness and the strength of the H$\alpha$ emission line was also observed during the first decrease of the V-band brightness. However, the H$\alpha$ did not display a very strong emission during the subsequent abrupt decrease of the optical brightness. With the extension of the decretion disk, it will be truncated by the gravitational interaction of the neutron star \citep{Okazaki01}. The efficiency of the truncation depends strongly on the gap between the truncation radius and the inner Lagrangian point and the viscosity of the disk. When the neutron star passes the periastron point, the decretion disk will be truncated efficiently. We can see from Figure~\ref{figure:hdpart2} that after the giant X-ray outbursts, the intensity of the H$\alpha$ decreased rapidly, which meant that the outer part of the decretion disk had been truncated by the neutron star. For the lines formed in the inner region of the decretion disk, such as HeI~$\lambda$6678, they could not truncated by the neutron star even around the periastron. Figure 1 in \citet{Haigh04} indicates that when the H$\alpha$ intensity showed a gradual decrease after the 1994 giant X-ray outburst, the HeI~$\lambda$6678 still had a strong emission. Our observational results shown in Figure~\ref{figure:spectra} also indicated that the HeI~$\lambda$6678 during our 2010 observations was still in a strong emission even the H$\alpha$ has decreased to its 2007 level.

A0535+26 underwent disk loss and subsequent renewal events during Cycle II in Figure~\ref{figure:hdpart1}. The anti-correlation between the optical brightness and the H$\alpha$ intensity was observed between MJD~50350 and MJD~50550. Due to bad coverage of the V-band magnitude and the H$\alpha$ EWs in the following 100 days, we could not know when the optical brightness decreased to its minimum and the intensity of H$\alpha$ reached its maximum. In Cycle II, both the optical and the H$\alpha$ line emission kept fading after MJD~50670. It seems that the decay timescales of the optical brightness are different during the increasing and decreasing phases of the H$\alpha$ intensity. They should have different physical origins. The optical fading between MJD~50350 and MJD~50550 might be caused by the mass ejection from the Oe star, and the subsequent fading of the optical brightness should be the indicator for the dilution of the decretion disk around the Oe star. The increase of the H$\alpha$ intensity at the beginning of Cycle II means the ejected mass had been transferred into the decretion disk. After the mass ejection event, a low-density region would be formed around the Oe star. With the extension of this low density region, the decretion disk will be a ring-like structure. If there was no mass ejection taking place in the system, the disk around the Oe star will be lost. Because the inner region of the disk has a major contribution to the optical continuum emission while the H$\alpha$ emission could be formed in the entire disk, the optical brightness of A0535+26 reached a stable faint state around MJD~50800 before the H$\alpha$ line turned from emission into absorption around MJD~51100. The subsequent disk renewal phase may be caused by the discrete mass ejections from the Oe star. Due to the smaller emission region for the optical continuum, the intensity of H$\alpha$ still kept increasing, when the V-band brightness showed a stable level during Cycle II.

The 2005 giant X-ray outburst in Cycle III of Figure~\ref{figure:hdpart2} was also observed during the fading phase of the optical brightness in A0535+26. The intensity of the H$\alpha$ line did not showed the obvious increase before the 2005 giant X-ray outburst. However, it shows an abrupt decrease after the giant X-ray outburst, which might be caused by the truncation of the neutron star. A pre-outburst structure was observed in the following normal  X-ray outburst around MJD~53600 \citep{Caballero08}, which might also be connected with the low-density region formed after the mass ejection.

\section{Conclusions}

The V-band optical lightcurve of A0535+26 indicates that each giant X-ray outburst occurred during a fading of the optical brightness in the system. The anti-correlation between the V-band magnitudes and the H$\alpha$ EWs during our 2009 observations indicated a mass ejection event had taken place before the 2009 giant X-ray outburst, which caused the formation of a low-density region in the disk. According to the diffusion time of the low-density region from the Oe star to the periastron point, the viscosity parameter is estimated to be $\sim$0.47. The rapid decrease of the H$\alpha$ emission after each giant X-ray outburst might be connected with the truncation of the outer disk by the neutron star in the system. The mass ejection occurred around 1996 September might trigger the subsequent disk loss event.

\acknowledgements
We thank the anonymous referee for insightful suggestions. JZY acknowledges Sylvain Chaty, E.P.J. van den Heuvel, and J.S., Clark for their useful discussions and Yizhong Fan his improvement of our manuscript. This work was supported by the National Basic Research Program of China - 973 Program 2009CB824800, the Natural Science Foundation of China under Grants 10873036 and 11003045, the National High Technology research and Development Program of China - 863 project 2008AA12Z304.


\begin{thebibliography}{}


\bibitem[Bjorkman \& Carciofi(2005)]{Bjorkman05}
Bjorkman, J.E., \& Carciofi, A.C., 2005, in ASP Conf. Ser. 337, The Nature and Evolution of Disks Around Hot Stars, ed. R. Ignace \& K. Gayley (San Francisco, CA: ASP), 75



\bibitem[Caballero et al.(2007)]{Caballero07}
Caballero, I., Kretschmar, P., Santangelo, A., et al., 2007, \aap, 465 L21

\bibitem[Caballero et al.(2008)]{Caballero08}
Caballero, I., Santangelo, A., Kretschmar, P., et al., 2008, \aap, 408, L17

\bibitem[Camero-Arranz et al.(2011)]{Camero-Arranz11}
Camero-Arranz, A., Finger, M.H., Wilson-Hodge, C.A., et al., 2011, arXiv:1109.3924
\bibitem[Clark et al.(1998)]{Clark98}
Clark, J. S., Tarasov, A. E., Steele, I. A., et al., 1998, MNRAS, 294, 165

\bibitem[Clark et al.(1999)]{Clark99}
Clark, J. S.; Lyuty, V. M.; Zaitseva, G. V., et al., 1999, MNRAS, 302, 167


\bibitem[Coe et al.(2006)]{Coe06}
Coe, M.J., Reig, P., McBride, V.A., et al., 2006, MNRAS, 368, 447

%


\bibitem[Finger et al.(1994)]{Finger94}
Finger, M. H., Wilson, R. B., \& Hagedon, K. S., 1994, IAUC, 5931

\bibitem[Finger et al.(1996)]{Finger96}
Finger, M.H., Wilson, R.B., \& Harmon, B.A., 1996, \apj, 459, 288

\bibitem[Giangrande et al.(1980)]{Giangrande80}
Giangrande, A., Giovannelli, F., Bartolini, C., et al., 1980, A\&AS, 40, 289

\bibitem[Giovannelli \& Graziati(1992)]{Giovannelli92}
Giovannelli, F. \& Graziati, L. S. 1992, Space Science Reviews, 59, 1


\bibitem[Grundstrom et al.(2007)]{Grundstrom07}
Grundstrom, E. D., Boyajian, T. S., Finch, C., et al., 2007, \apj, 660, 1398

\bibitem[Haigh et al.(1999)]{Haigh99}
Haigh, N. J., Coe, M. J., Steele, I. A., \& Fabregat, J., 1999, MNRAS, 311, L21


\bibitem[Haigh et al.(2004)]{Haigh04}
Haigh, N. J., Coe, M. J., \& Fabregat, J., 2004, MNRAS, 350, 1457

\bibitem[Hao et al.(1996)]{Hao96}

Hao, J.-X., Huang, L., \& Guo, Z. H., 1996, \aap, 308, 499

\bibitem[Janot-Pacheco et al.(1987)]{Janot-Pacheco87}

Janot-Pacheco, E., Motch, C., \& Mouchet, M., 1987, \aap, 177, 91

\bibitem[Liu et al.(2006)]{Liu06}
Liu, Q. Z., van Paradijs, J., and van den Heuvel, E. P. J., 2006, \aap, 455, 1165

\bibitem[Liu et al.(2011)]{Liu11}
Liu, Q.Z., Li, H., Yan, J.Z., 2011, ApJ to be submitted

\bibitem[Lyuty \& Zaitseva(2000)]{Lyuty00}
Lyuty, V. M. \& Zaitseva, G. V., 2000, Astronomy Letters, 26, 9



\bibitem[Makino et al.(1989)]{Makino89}
Makino, F., Cook, W., Grunsfeld, J., et al., 1989, IAUC, 4769

\bibitem[Meilland et al.(2006)]{Meilland06}
Meilland, A., Stee, Ph., Zorec, J., and Kanaan, S., 2006, \aap, 455, 953

\bibitem[Moritani et al.(2010)]{Moritani2010}
Moritani, Y., Nogami, D., Okazaki, A. T., et al., 2010, MNRAS, 405, 467

\bibitem[Nagase et al.(1982)]{Nagase82}
Nagase, F., Hayakawa, S., Kunieda, H., et al., 1982, \apj, 263, 814

\bibitem[Okazaki \& Negueruela(2001)]{Okazaki01}
Okazaki, A. T. and Negueruela, I., 2001, \aap, 377, 161


\bibitem[Porter \& Rivinius(2003) ]{Porter03}
Porter, J. M. and Rivinius, T., 2003, PASP, 115, 1153


\bibitem[Reig(2011)]{Reig11}
Reig, P., 2011, Astrophysics and Space Science, 332, 1

\bibitem[Reynolds \& Miller(2010)]{Reynolds2010}
Reynolds, M.T., and Miller, J.M., 2010, \apj, 723,1799

\bibitem[Rivinius et al.(2001)]{Rivinius01}
Rivinius, Th., Baade, D., Stefl, S., and Maintz, M., 2001, \aap, 379, 257

\bibitem[Rosenberg et al.(1975)]{Rosenberg75}
Rosenberg, F. D., Eyles, C. J., Skinner, G. K., \& Willmore, A. P.,  1975, Nature,256, 628

\bibitem[Sembay et al.(1990)]{Sembay90}
Sembay, S., Schwartz, R. A., Orwig, L. E., et al., 1990, \apj, 351, 675

\bibitem[Shakura \& Syunyaev(1973)]{Shakura73}
Shakura, N.I., \& Syunyaev, R.A., 1973, \aap, 24, 337
\bibitem[Slettebak et al.(1992)]{Slettebak92}
Slettebak, A., Collins, G.W., II, Truax, R., 1992, ApJS, 81, 335


\bibitem[Stee et al.(1998)]{Stee98}
Stee, Ph., Vakili, D., Bonneau, D., \& Mourard, D., 1998, \aap, 332, 268
\bibitem[Tueller et al.(2005)]{Tueller05}
Tueller, Jack, Ajello, Marco, \& Barthelmy, Scott, 2005, ATel, 504


\bibitem[Wilson-Hodge et al.(2009)]{Wilson-Hodge09}
Wilson-Hodge, Colleen A.,  Finger, Mark H, Camero-Arranz, Ascension, et al., 2009, ATel, 2324

\bibitem[Zaitseva(2005)]{Zaitseva05}
Zaitseva, G.V., 2005, Astronomy Letters, 31, 103





\end{thebibliography}
\end{document}